\documentclass[onecollarge, natbib]{svjour2}
\bibpunct{[}{]}{;}{n}{}{,} % to get "[numbered]" references from natbib
\smartqed  % flush right qed marks, e.g. at end of proof
\usepackage{graphicx}
\usepackage{latexsym}
\usepackage{amsmath}
\usepackage{epstopdf}
\usepackage{subfigure}
\journalname{Few-Body Systems}
\begin{document}

\title{Exactly solvable dynamical models with a minimal length uncertainty %\thanks{Grants or other notes
}


\author{Reginald Christian S. Bernardo         \and
        Jose Perico H. Esguerra %etc.
}


\institute{R. C. S. Bernardo \at
              Theoretical Physics Group, National Institute of Physics, University of the Philippines, Diliman, Quezon City 1101, Philippines \\
              \email{rcbernardo@nip.upd.edu.ph}           %  \\
           \and
           J. P. H. Esguerra \at
              Theoretical Physics Group, National Institute of Physics, University of the Philippines, Diliman, Quezon City 1101, Philippines \\
              \email{perry.esguerra@gmail.com}
}

\date{Received: date / Accepted: date}

\maketitle

\begin{abstract}
We present exact analytical solutions to the classical equations of motion and analyze the dynamical consequences of the existence of a minimal length for the free particle, particle in a linear potential, anti-symmetric constant force oscillator, harmonic oscillator, vertical harmonic oscillator, linear diatomic chain, and linear triatomic chain. It turns out that a minimal length increases the speed of a free particle and the rate of fall of a particle that is subject to the influence of a linear potential. Our results suggest that the characteristic frequency of systems tend to increase when there is a minimal length. This is a common feature that we observed for the oscillator systems that we have considered. 
 
\end{abstract}

%---------------------------------------------------------------------------
%               main text
%---------------------------------------------------------------------------
\section{Introduction}
\label{sec:intro}

In 1995, Kempf, Mangano, and Mann \cite{hilbert_kmm} proposed a modification of quantum mechanics that phenomenologically includes a minimal uncertainty in position or minimal length through the generalized uncertainty principle (GUP) given by
\begin{equation}
\label{eq:gup}
\Delta X \Delta P \geq \frac{\hbar}{2} ( 1 + \beta ( \Delta P )^2 + \zeta)
\end{equation}
where $X$ is the position, $P$ is the momentum, $\beta$ is the GUP parameter, and $\zeta$ is a positive constant that depends on the expectation value of the momentum. It has been suggested that a theory with the GUP can be used to describe nonpointlike particles such as molecules and nucleons \cite{quasiparticles_kempf, quasiparticles_sastry} although interest in the GUP comes mostly from the fact that it arises from string theory \cite{string_konishi1, string_maggiore1, string_maggiore2, string_maggiore3}
and that quantum gravity theories \cite{q_geom_capozziello1, black_hole_physics_scardigli1} predict the existence of a minimal length. Phenomenological investigation of quantum gravity effects have therefore been considered using the GUP \cite{hydrogen_brau2, hydrogen_bouaziz1, harmonic_oscillator_pedram6, harmonic_chang1, quantum_bouncer_pedram1, quantum_bouncer_brau1, scattering_hassanabadi1, scattering_haouat1, cusp_hassanabadi}. It is worthwhile to consider the dynamical implications of the existence of a minimal length on systems because of suggestions \cite{dynamics_hassanabadi, dynamics_nozari1, dynamics_nozari2} that the Ehrenfest theorem is not valid when the GUP is considered. These suggestions come from comparing quantum mechanics with the GUP with the usual classical dynamics and therefore need to be given further attention by considering a GUP modified classical dynamics. There have been relatively few investigations on the dynamical implications of the GUP \cite{dynamics_hassanabadi, composite_tkachuk1, dynamics_nozari1} and a complete picture of the theory requires an analysis of time development. 

We address this issue in this paper and start the analysis of time development of classical systems with a minimal length. This is a reasonable first approach to dynamics;  while an analysis of quantum dynamics with the GUP often leads to a challenging set of operator differential equations, it turns out that some analytical progress can be made on the counterpart classical dynamics. We hope that the insight that can be gained from studying the classical counterpart can provide hints about the quantum dynamics. We begin by writing the equations of motion for the classical system when there is a minimal length (Section \ref{sec:dynamics_gup}). We then present exact analytical solutions to the equations of motion and analyze the classical dynamical implications of a minimal length for the free particle, particle in a linear potential, anti-symmetric constant force oscillator, harmonic oscillator, vertical harmonic oscillator, linear diatomic chain, and linear triatomic chain (Section \ref{sec:exact_models}).

\section{Classical dynamics with minimal length}
\label{sec:dynamics_gup}

The GUP given by Eq. \ref{eq:gup} predicts the existence of a minimal uncertainty in position or a minimal length of magnitude $\Delta X_{min} = \hbar \sqrt{ \beta }$. One way of arriving at this GUP is to consider the deformed commutation relation given by \cite{hilbert_kmm}
\begin{equation}
\label{eq:dcr}
[\hat{X}, \hat{P}] = i \hbar ( 1 + \beta \hat{P}^2 ) .
\end{equation}
The classical counterpart of Eq. \ref{eq:dcr} is the deformed Poisson bracket given by \cite{orbits_benczik1, orbits_silagadze1, composite_tkachuk1} 
\begin{equation}
\label{eq:dpb} \{X, P \} = 1 + \beta P^2 .
\end{equation}
We can show that this deformed Poisson bracket is satisfied by the classical dynamical position $X$ and momentum $P$ given by
\begin{eqnarray}
\label{eq:X} X &=& x \\
\label{eq:P} P &=& p + \frac{1}{3} \beta p^3
\end{eqnarray}
where $x$ and $p$ which we will refer to as the pseudoposition and pseudomomentum, respectively, satisfy $\{x, p \} = 1$. The Hamiltonian is given by
\begin{equation}
\label{eq:hamiltonian}
H = \frac{P^2}{2m} + V(X) = \frac{p^2}{2m} + \frac{1}{3} \frac{\beta}{m} p^4 + V(x) .
\end{equation}
It follows from $X = x$ and the principle of conservation of energy that the kinetic energy of the system can be measured once the functional form of the potential acting on the system is known. Because the fundamental form of the kinetic energy is written as the single term $\frac{P^2}{2m}$ it follows that $P$ is what we know and measure to be the momentum. But although $X$ and $P$ are the physical quantities it turns out that $x$ and $p$ are useful mathematical tools and that we can analyze the dynamics of the system by studying the time evolution of $x$ and $p$. 

Using the Hamiltonian equations of motion, we can show that the pseudoposition $x$ and pseudomomentum $p$ evolve in time according to
\begin{equation}
\label{eq:velocity}
\frac{d x}{dt} = \frac{p}{m} + \frac{4}{3} \frac{\beta}{m} p^3
\end{equation}
\begin{equation}
\label{eq:force}
\frac{d p}{dt} = - \frac{d V}{dx} .
\end{equation}
Complete dynamical information can by obtained by solving Eqs. \ref{eq:velocity} and \ref{eq:force}  for $x$ and $p$ subject to the initial conditions
\begin{eqnarray}
x|_{t = 0} &=& x_0 \\
p|_{t = 0} &=& p_0 .
\end{eqnarray}
In the final analysis we want the physical quantities $X$ and $P$ to be expressed entirely in terms of the initial quantities $X_0$ and $P_0$. In this case we plug in our solutions to $x$ and $p$ back into Eqs. \ref{eq:X} and \ref{eq:P} and use the relations
\begin{equation}
\label{eq:x_0_X_0}
x_0 = X_0
\end{equation}
and 
\begin{equation}
\label{eq:p_0_P_0}
p_0 = - \bigg(  \frac{2}{ 3 \beta^2 P_0 + \sqrt{ 4 \beta^3 + 9 \beta^4 P_0^2   }     }     \bigg)^{\frac{1}{3}}    +  \bigg(   \frac{3 \beta^2 P_0 + \sqrt{4 \beta^2 + 9 \beta^4 P_0^2}  }{2 \beta^3 }       \bigg)^{\frac{1}{3}} .
\end{equation}
Eqs. \ref{eq:x_0_X_0} and \ref{eq:p_0_P_0} are the inverse of Eqs. \ref{eq:X} and \ref{eq:P} at $t = 0$.

\section{Exactly solvable models}
\label{sec:exact_models}

\subsection{Free particle}
\label{sec:free_particles}

For the free particle $V(x) = V_0$. Eq. \ref{eq:force} gives $\frac{dp}{dt} = 0$ and hence the pseudomomentum is a constant of motion
\begin{equation}
\label{eq:p_free_particle}
p(t) = p(t = t_0) = p_0 .
\end{equation}
Plugging in Eq. \ref{eq:p_free_particle} into Eq. \ref{eq:velocity} and integrating with respect to time $t$ we obtain the pseudoposition
\begin{equation}
\label{eq:x_free_particle}
x(t) = x_0 + \bigg( \frac{p_0}{m} + \frac{4}{3} \frac{\beta}{m} p_0^3 \bigg) t . 
\end{equation}
This result shows that the existence of a minimal length tends to increase the velocity of a free particle.

\subsection{Particle in a linear potential}
\label{sec:uniform_field}

In this case we have $V(x) = b x$. The constant $b$ is the weight for a massive particle in a uniform gravitational field and the electric force for a charged particle in a uniform electric field. Using Eq. \ref{eq:force} we have
\begin{equation}
\label{eq:force_linear_potential}
\frac{dp}{dt} = -b .
\end{equation}
Eq. \ref{eq:force_linear_potential} can be readily integrated for the pseudomomentum
\begin{equation}
\label{eq:p_linear_potential}
p(t) = -b t + p_0 .
\end{equation}
Plugging in Eq. \ref{eq:p_linear_potential} into Eq. \ref{eq:velocity} gives
\begin{equation}
\label{eq:v_linear_potential}
\frac{dx}{dt} = \frac{-bt + p_0}{m} + \frac{4}{3} \frac{\beta}{m} (-bt + p_0)^3 .
\end{equation}
The integration of Eq. \ref{eq:v_linear_potential} with respect to the time $t$ is straightforward and gives the pseudoposition
\begin{equation}
\label{eq:x_linear_potential}
x(t) = x_0 + \frac{p_0}{m} t  - \frac{b}{2m} t^2 - \frac{1}{3} \frac{\beta}{m b} [( -bt + p_0 )^4 - p_0^4].
\end{equation}
Eq. \ref{eq:x_linear_potential} shows that the effect of a minimal observable length on a particle that is subject to the influence of a constant force is to increase the rate of fall to the region of lower potential.

\subsection{Anti-symmetric constant force oscillator}
\label{sec:abs_val_oscillator}

We consider a particle bound by the absolute value potential given by $V(x) = b|x|$. Using Eq. \ref{eq:force} this gives rise to the anti-symmetric constant force oscillator
\begin{equation}
\label{eq:force_sgn}
\frac{dp}{dt} = -b \ \textrm{sgn}(x)
\end{equation}
where $\textrm{sgn}(x)$ is the the signum function. 
Before a quarter of a period $T$ we can integrate the pseudoposition $x$ and pseudomomentum $p$ from Eqs. \ref{eq:force_sgn} and \ref{eq:velocity} with the particular initial conditions $x_0 > 0$ and $p_0 = 0$. The results are
\begin{equation}
\label{eq:p_sgn}
p(t) = -bt
\end{equation}
\begin{equation}
\label{eq:x_sgn}
x(t) = x_0 - \frac{bt}{2m} - \frac{\beta b^3 t^4}{3m}.
\end{equation}
The period of motion can be determined by setting $x = 0$ at $t = \frac{T}{4}$. This leads to
\begin{equation}
\label{eq:period_sgn}
\frac{T}{4} = \sqrt{\frac{3}{4 \beta b^2}} \sqrt{  \sqrt{1 + \frac{16 \beta b x_0 m}{3}} -1  } .
\end{equation}
Using the principle of conservation of energy we can check that the pseudoposition and pseudomomentum have the same period. Using Eq. \ref{eq:hamiltonian} we obtain
\begin{equation}
\label{eq:energy_conservation_sgn}
bx_0 = \frac{p_f^2}{2m} + \frac{1}{3} \frac{\beta}{m} p_f^4 
\end{equation}
where $p = p_f$ when $x = 0$. It is straightforward to solve for $p_f$ in Eq. \ref{eq:energy_conservation_sgn}
\begin{equation}
\label{eq:p_f_sgn}
p_f =  \sqrt{ \frac{3}{4 \beta} } \sqrt{ \sqrt{1 + \frac{16 \beta b x_0 m}{3}} -1 } .
\end{equation}
Using Eqs. \ref{eq:period_sgn} and \ref{eq:p_f_sgn} we can check that $p(t = \frac{T}{4}) = p_f$. Therefore it is confirmed that pseudoposition and pseudomomentum have the same period. Correspondingly, the position and momentum (Eqs. \ref{eq:X} and \ref{eq:P}) will have the same period.

\subsection{Harmonic oscillator}
\label{sec:harmonic_oscillator}

\begin{figure}[h!]
\center
	\subfigure[]{
		\includegraphics[width = 4.0in]{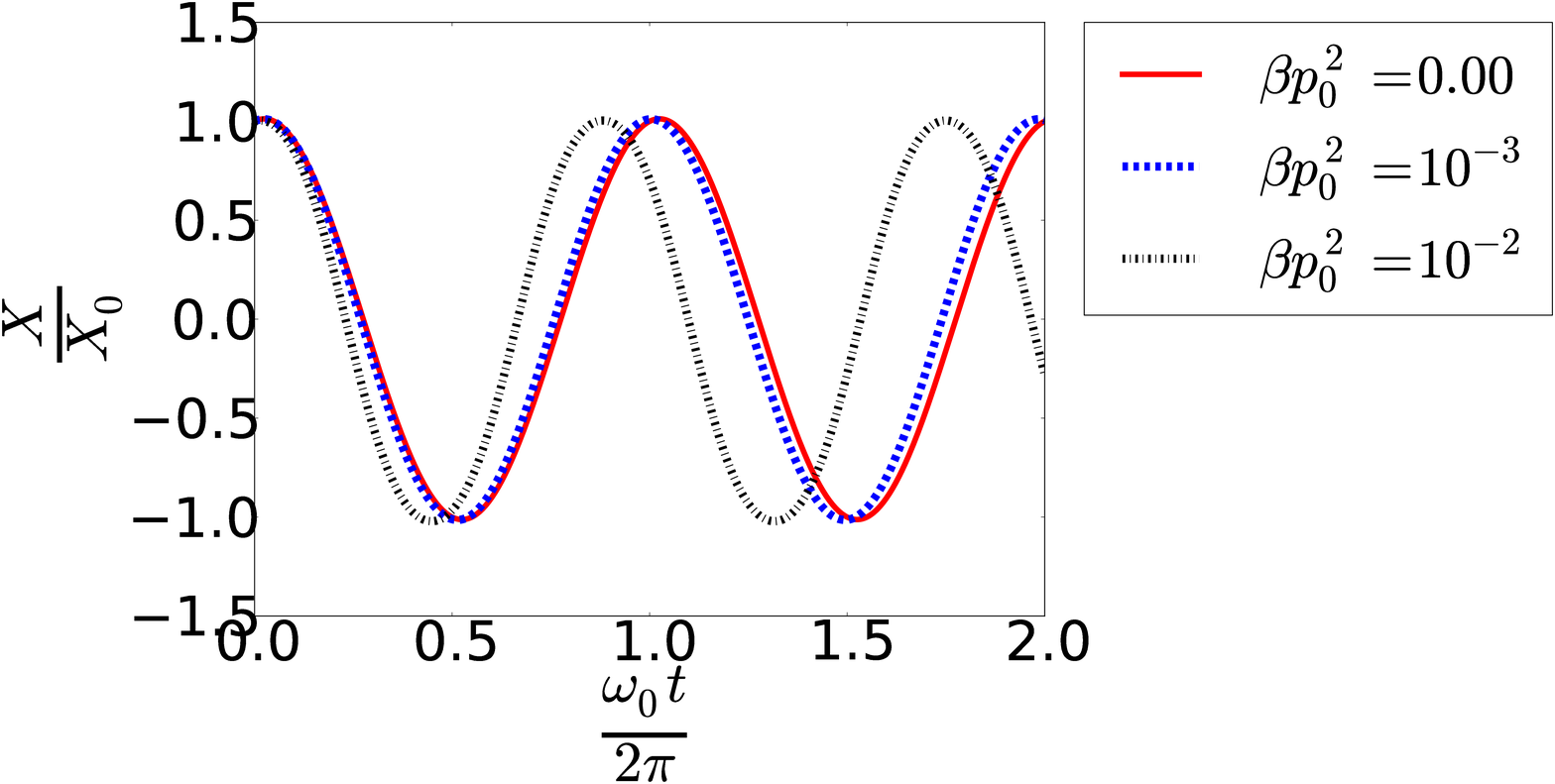}
		}
	\subfigure[]{
		\includegraphics[width = 4.0in]{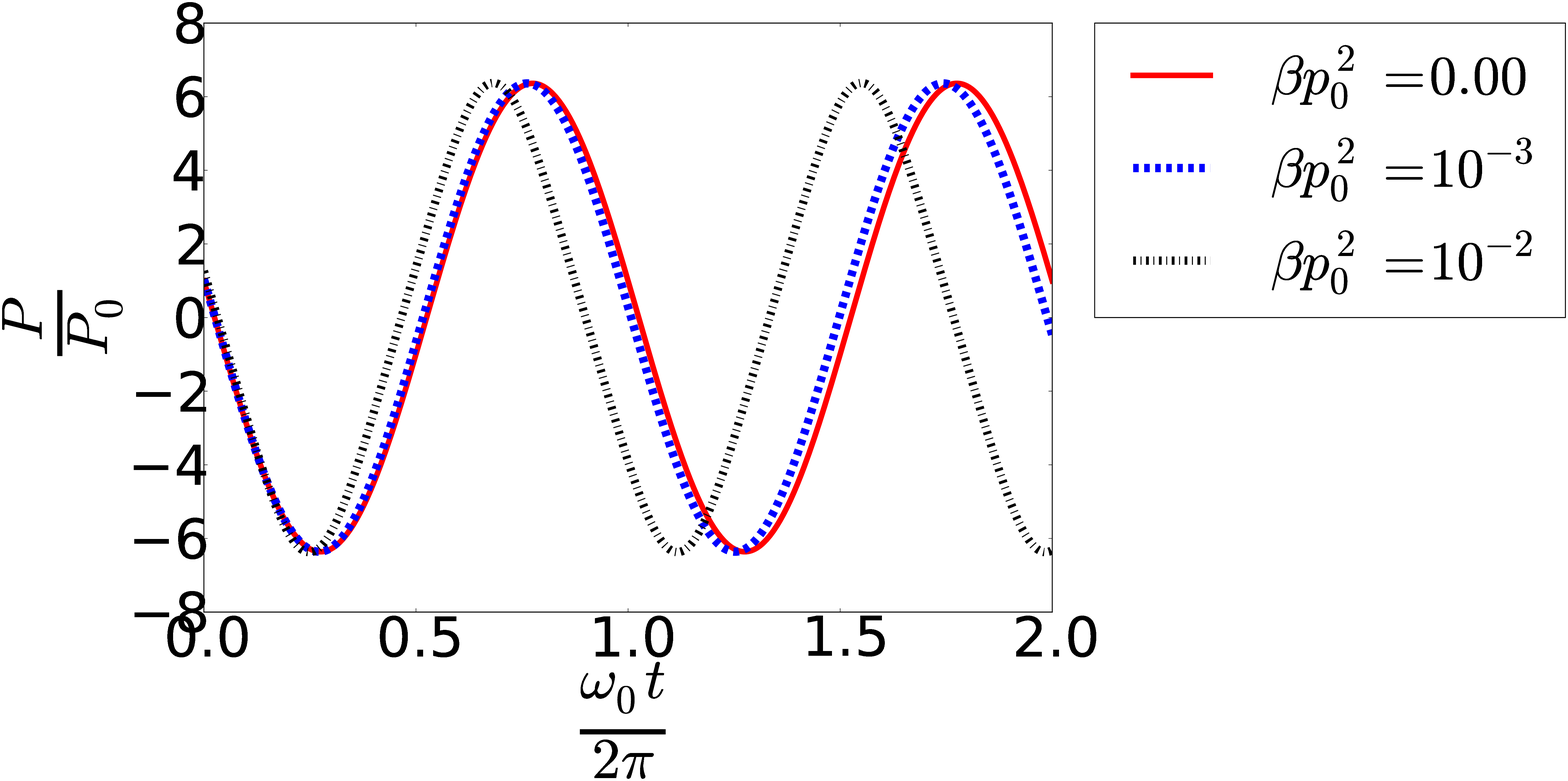}
		}
	\subfigure[]{
		\includegraphics[width = 4.0in]{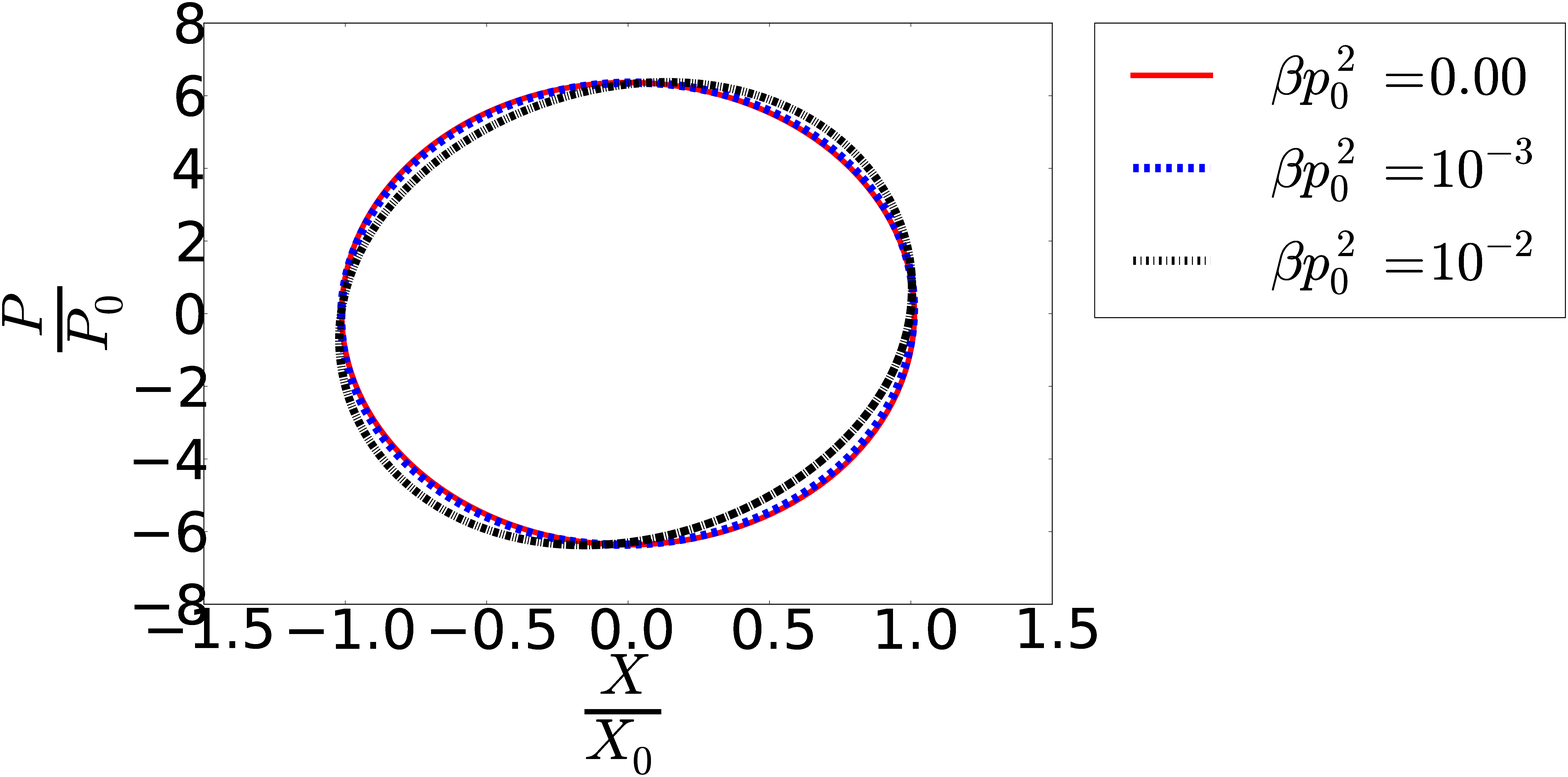}
	}
\caption{(a) Position and (b) momentum as a function of time for two ordinary cycles with corresponding (c) phase space for a harmonic oscillator of unit mass.}
\label{fig:harmonic_oscillator}
\end{figure}

For the harmonic oscillator $V(x) = \frac{1}{2} m \omega^2 x^2$. Eq. \ref{eq:force} gives
\begin{equation}
\label{eq:force_ho}
\frac{dp}{dt} = - m\omega_0^2 x .
\end{equation}
We can uncouple $p$ by differentiating Eq. \ref{eq:force_ho} and using Eq. \ref{eq:velocity} to eliminate $x$. This leads to the Duffing equation \cite{duffing_amore1, duffing_belendez1}
\begin{equation}
\label{eq:momentum_second_derivative}
\frac{d^2p}{dt^2} + \omega_0^2 p + \frac{4}{3} \omega_0^2 \beta p^3 = 0 .
\end{equation}
This equation was also encountered by Hassanabadi, Hooshmand, and Zarrinkamar \cite{dynamics_hassanabadi} in their analysis of the same problem. In this work, we present an exact solution to the Duffing equation in terms of elliptic functions. The solution to Eq. \ref{eq:momentum_second_derivative} subject to the initial conditions
\begin{equation}
\label{eq:initial_conditions_momentum}
p|_{t = 0} = p_0 \ \ \ \ \ \ \frac{dp}{dt} \bigg|_{t = 0} = -m \omega_0^2 x_0
\end{equation}
is given by \cite{duffing_belendez1}
\begin{equation}
\label{eq:momentum_ho}
p(t) = A \ \textrm{cn} ( \Omega (t - t_1), \mu )
\end{equation}
where
\begin{equation}
\label{eq:frequency_effective_ho}
\Omega = \omega_0 \sqrt{1 + \frac{4}{3} \beta A^2}
\end{equation}
\begin{equation}
\label{eq:momentum_amp}
A = \sqrt{\frac{3}{4 \beta}}  \sqrt{ \sqrt{1 + \frac{8 \beta }{3} \bigg( m^2 \omega_0^2 x_0^2 + p_0^2 + \frac{2}{3} \beta p_0^4   \bigg)}   - 1}
\end{equation}
\begin{equation}
\label{eq:momentum_modulus}
\mu = \frac{2}{3} \beta \bigg(  \frac{\omega_0 A}{\Omega}  \bigg)^2 
\end{equation}
\begin{equation}
\label{eq_momentum_t_1}
t_1 = \frac{1}{ \Omega } \bigg( \textrm{K} \bigg( \frac{p_0}{A}, \mu   \bigg) + \frac{\pi}{2} \bigg)
\end{equation}
and $\textrm{cn}(x, \mu)$ and $\textrm{K}(x, \mu)$ are the Jacobi elliptic function and incomplete elliptic integral of the first kind, respectively, with modulus $\mu$ \cite{jacobi_baker1, jacobi_cervero1, jacobi_lawden1}. The pseudomomentum $p$ has a well-defined frequency given by
\begin{equation}
\label{eq:momentum_frequency}
\omega = \frac{\pi}{2} \frac{\Omega}{\textrm{K}(\mu)}
\end{equation}
where $\textrm{K}(m)$ is the complete elliptic integral of the first kind \cite{jacobi_lawden1}. We can determine the pseudoposition $x$ by plugging in Eq. \ref{eq:momentum_ho} into Eq. \ref{eq:velocity} and integrating over the time $t$. It is easy to show that 
\begin{equation}
\label{eq:integ_p}
\int_0^t dt' p(t') = \frac{A}{\Omega} \frac{ \textrm{z}(t, \mu)}{\mu}
\end{equation}
\begin{equation}
\label{eq:integ_p_cube}
\int_0^t dt' p^3(t') = \frac{A^3}{\Omega} \frac{  (2 \mu^2 - 1) \textrm{z}(t, \mu) + \mu \textrm{y}(t, \mu))      }{ 2 \mu^3}
\end{equation}
where
\begin{equation}
\label{eq:z_t}
\textrm{z}(t, \mu) = \arcsin \bigg( \frac{2 \mu \ \textrm{sn}(\frac{\Omega t}{2}, \mu) \ \textrm{dn} (\frac{\Omega t}{2}, \mu) \ \textrm{cn} (\Omega (\frac{t}{2} - t_1), \mu )  }  {1 - \mu^2 \textrm{sn}^2 (\frac{\Omega t}{2}, \mu ) \textrm{sn}^2 (\Omega (\frac{t}{2} - t_1 ), \mu ) } \bigg)
\end{equation}
\begin{equation}
\label{eq:y_t}
y(t, \mu) = \textrm{sn} (\Omega (t - t_1), \mu) \ \textrm{dn} (\Omega (t - t_1), \mu) + \textrm{sn} (\Omega t_1, \mu) \ \textrm{dn} (\Omega t_1, \mu)
\end{equation}
and $\textrm{sn}(x, \mu)$ and $\textrm{dn}(x, \mu)$ are the Jacobi elliptic functions \cite{jacobi_baker1, jacobi_cervero1, jacobi_lawden1}. The pseudoposition is therefore given by
\begin{equation}
\label{eq:position_ho}
x(t) = x_0 + \frac{1}{m} \frac{A}{ \Omega } \frac{ \textrm{z}(t;\mu)}{\mu} + \frac{4}{3} \frac{\beta}{m } \frac{A^3}{\Omega} \frac{  (2 \mu^2 - 1) \textrm{z}(t; \mu) + \mu \textrm{y}(t; \mu))      }{ 2 \mu^3} .
\end{equation}
Fig. \ref{fig:harmonic_oscillator} shows a plot of the position and momentum (Eqs. \ref{eq:X} and \ref{eq:P}) and corresponding phase space for a harmonic oscillator of unit mass for three values of the GUP parameter $\beta$. 
Fig. \ref{fig:harmonic_oscillator} shows that the frequency of the harmonic oscillator increases when there is a minimal length.

\subsection{Vertical harmonic oscillator}
\label{sec:vertical_harmonic_oscillator}

Consider the vertical harmonic oscillator, classically representing a particle hung from a ceiling by a spring, with the potential $V(x) = \frac{1}{2}m \omega_0^2 (x - \xi)^2 + mgx $ . Eq. \ref{eq:force} leads to 
\begin{equation}
\label{eq:force_vertical_ho}
\frac{dp}{dt} = - m \omega_0^2 (x - \xi) - mg .
\end{equation}
We can uncouple equation Eqs. \ref{eq:force_vertical_ho} and \ref{eq:velocity} by differentiating Eq. \ref{eq:force_vertical_ho} with respect to the time $t$ and using Eq. \ref{eq:velocity}. This leads again to the Duffing equation exactly given by Eq. \ref{eq:momentum_second_derivative}. The only difference in this case is through the initial conditions
\begin{equation}
\label{eq:initial_conditions_vertical}
p|_{t = 0} = p_0 \ \ \ \ \ \ \frac{dp}{dt}\bigg|_{t = 0} = -m \omega_0^2 (x_0 - \xi) - mg .
\end{equation}
Hence we can simply consider the case of the vertical harmonic oscillator as a shifted harmonic oscillator and use the results of the previous subsection.

\subsection{Linear diatomic chain (two-body problem)}
\label{sec:diatomic_chain}

In this section and the next, we will consider two- and three-body problems which require us to make the assumption that the kinetic energies are additive. This assumption has already been used in a previous work and has also led to saving the equivalence principle \cite{composite_tkachuk1}. Moreover, the natural generalization to Eq. \ref{eq:dcr} for an $N$ body system is that the coordinates and momenta of different particles commute \cite{composite_buisseret1, composite_quesne1}. Thus, for the case of the linear diatomic chain we obtain the Hamiltonian
\begin{equation}
\label{eq:hamiltonian_diatomic_chain}
H = \frac{p_1^2}{2m_1} + \frac{1}{3} \frac{\beta_1}{m_1} p_1^4 + \frac{p_2^2}{2 m_2} + \frac{1}{3} \frac{\beta_2}{m_2} p_2^4 + \frac{1}{2}k (x_1 - x_2)^2 
\end{equation} 
where $x_1$, $p_1$, $m_1$, and $\beta_1$ and $x_2$, $p_2$, $m_2$, and $\beta_2$ are the pseudoposition, pseudomomenta, masses, and GUP parameters of particles 1 and 2, respectively. Note that we need not have the same GUP parameter $\beta$, unless we are dealing with the same elementary particle, but we require that
\begin{equation}
\label{eq:gamma}
\sqrt{\beta_1} m_1 = \sqrt{\beta_2} m_2 = \gamma
\end{equation}
where $\gamma$ is a universal constant. This ensures that the weak equivalence principle is satisfied \cite{composite_tkachuk1}.

The Hamiltonian equations of motion for the variables $x_1$, $p_1$, $x_2$, and $p_2$ are given by
\begin{equation}
\label{eq:vel_1_diatomic}
\frac{dx_1}{dt} = \frac{p_1}{m_1} + \frac{4}{3} \frac{\beta_1}{m_1} p_1^3
\end{equation}
\begin{equation}
\label{eq:force_1_diatomic}
\frac{d p_1}{dt} = - k( x_1 - x_2 )
\end{equation}
\begin{equation}
\label{eq:vel_2_diatomic}
\frac{dx_2}{dt} = \frac{p_2}{m_2} + \frac{4}{3} \frac{\beta_2}{m_2} p_2^3
\end{equation}
\begin{equation}
\label{eq:force_2_diatomic}
\frac{d p_2}{dt} = - k( x_2 - x_1 ) .
\end{equation}
Complete dynamical behaviour will be known once Eqs. \ref{eq:vel_1_diatomic}, \ref{eq:force_1_diatomic}, \ref{eq:vel_2_diatomic}, and \ref{eq:force_2_diatomic} have been solved for $x_1$, $p_1$, $x_2$ and $p_2$ subject to the initial conditions
\begin{equation}
\label{eq:initial_conditions_diatomic}
\begin{cases}  x_1|_{t = 0} = x_{10} \ \ \ \ \ \ p_1|_{t = 0} = p_{10}  \\ x_2|_{t = 0} = x_{20} \ \ \ \ \ \ p_2|_{t = 0} = p_{20} \end{cases}.
\end{equation}
By adding Eqs. \ref{eq:force_1_diatomic} and \ref{eq:force_2_diatomic} we obtain $\frac{d}{dt} (p_1 + p_2) = 0$. The total pseudomomentum $p_1 + p_2$ is therefore a constant of motion
\begin{equation}
\label{eq:total_momentum_diatomic}
p_T = p_1 + p_2 .
\end{equation}
We can use Eq. \ref{eq:total_momentum_diatomic} to eliminate either $p_1$ or $p_2$ in consideration. Eliminating $p_2$ from Eq. \ref{eq:vel_2_diatomic} yields
\begin{equation}
\label{eq:vel_2_diatomic_p_1}
\frac{dx_2}{dt} = \frac{p_T - p_1}{m_2} + \frac{4}{3} \frac{\beta_2}{m_2} (p_T - p_1)^3 .
\end{equation}
Differentiating Eq. \ref{eq:force_1_diatomic} and using Eqs. \ref{eq:vel_1_diatomic} and \ref{eq:vel_2_diatomic_p_1} yield
\begin{equation}
\label{eq:p_1_second_derivative}
\frac{d^2 p_1}{dt^2} = -k \bigg( \frac{p_1}{m_1} + \frac{4}{3} \frac{\beta_1}{m_1} p_1^3 \bigg) + k \bigg[ \frac{(p_T - p_1)}{m_2} + \frac{4}{3} \frac{\beta_2}{m_2} (p_T - p_1)^3  \bigg] .
\end{equation}

We now specialize to the frame where $p_T = 0$. In this case Eq. \ref{eq:p_1_second_derivative} becomes the Duffing equation
\begin{equation}
\label{eq:duffing_eq_diatomic}
\frac{d^2 p_1}{dt^2} + \frac{k}{m^\ast} p_1 + \frac{4}{3} k \frac{\beta ^\ast}{m^\ast}  p_1^3 = 0 
\end{equation}
where 
\begin{equation}
\label{eq:mass_reduced}
\frac{1}{m^\ast} = \frac{1}{m_1} + \frac{1}{m_2}
\end{equation}
and 
\begin{equation}
\label{eq:gup_reduced}
\beta^\ast = \frac{m_1 \beta_2 + m_2 \beta_1 }{m_1 + m_2} .
\end{equation}
We know that we can solve Eq. \ref{eq:duffing_eq_diatomic} for the pseudomomentum $p_1$ with the initial conditions
\begin{equation}
\label{eq:initial_conditions_p_1_diatomic}
p_1|_{t = 0} = p_{10} \ \ \ \ \ \ \frac{dp_1}{dt} \bigg|_{t = 0} = -k(x_{10} - x_{20}) 
\end{equation}
in terms of a Jacobi elliptic function (Eq. \ref{eq:momentum_ho}). The situation is therefore similar to the harmonic oscillator but with a reduced mass $m^\ast$ and reduced GUP parameter $\beta^\ast$. Our expression for the reduced GUP parameter (Eq. \ref{eq:gup_reduced}) agrees with Quesne and Tkachuk's result for the effective deformation parameter for relative motion in a two-body problem \cite{composite_quesne1}. We can plug back the pseudomomentum $p_1$ to Eq. \ref{eq:vel_1_diatomic} and \ref{eq:vel_2_diatomic_p_1} and integrate exactly to obtain the pseudopositions $x_1$ and $x_2$. The pseudomomentum $p_2$ can be obtained from Eq. \ref{eq:total_momentum_diatomic}. We can solve for a center of mass velocity by multiplying Eqs. \ref{eq:vel_1_diatomic} and \ref{eq:vel_2_diatomic} by $m_1$ and $m_2$, respectively, noting that we have specialized to the frame in which $p_T = 0$, and adding the resulting equations. This gives
\begin{equation}
\label{eq:center_of_mass_velocity_diatomic}
m_1 \frac{d x_1}{dt} + m_2 \frac{d x_2}{dt} = \frac{4}{3} (\beta_1 - \beta_2) p_1^3 .
\end{equation}
As a consequence of the existence of a minimal length we see that the center of mass is not at rest but oscillates with a well-defined frequency in the frame in which $p_T = 0$. A special case arises when the diatomic chain is made up of the same elementary particles ($\beta_1 = \beta_2$ and $m_1 = m_2$). In this case, the center of mass is at rest. Using Eq. \ref{eq:integ_p_cube} to integrate Eq. \ref{eq:center_of_mass_velocity_diatomic} we can obtain an exact-closed form expression for the center of mass coordinate ($R = m_1 X_1 + m_2 X_2$).

\subsection{Linear triatomic chain (three-body problem)}
\label{sec:triatomic chain}

Finally, we study the effect of minimal length on the longitudinal modes of the symmetrical linear triatomic chain. In general, the Hamiltonian for the linear triatomic chain is
\begin{equation}
\label{eq:hamiltonian_triatomic}
H = \frac{p_1^2}{2m_1} + \frac{1}{3} \frac{\beta_1}{m_1} p_1^4 + \frac{p_2^2}{2m_2} + \frac{1}{3} \frac{\beta_2}{m_2} p_2^4 + \frac{p_3^2}{2m_3} + \frac{1}{3} \frac{\beta_3}{m_3} p_3^4  + \frac{1}{2} k_1 (x_1 - x_2)^2 + \frac{1}{2} k_3 (x_3 - x_2)^2 
\end{equation}
where $x_i$, $p_i$, $m_i$, and $\beta_i$ are the pseudopositions, pseudomomenta, masses, and the GUP parameters, respectively, of the ith particle. Using Hamiltonian equations of motion on each of the $x_i$ and $p_i$ yields the set of differential equations
\begin{equation}
\label{eq:vel_1_triatomic}
\frac{dx_1}{dt} = \frac{p_1}{m_1} + \frac{4}{3} \frac{\beta_1}{m_1} p_1^3
\end{equation}
\begin{equation}
\label{eq:force_1_triatomic}
\frac{d p_1}{dt} = -k_1 (x_1 - x_2)
\end{equation}
\begin{equation}
\label{eq:vel_2_triatomic}
\frac{dx_2}{dt} = \frac{p_2}{m_2} + \frac{4}{3} \frac{\beta_2}{m_2} p_2^3
\end{equation}
\begin{equation}
\label{eq:force_2_triatomic}
\frac{d p_2}{dt} = -k_1 (x_2 - x_1) - k_3 (x_2 - x_3)
\end{equation}
\begin{equation}
\label{eq:vel_3_triatomic}
\frac{dx_3}{dt} = \frac{p_3}{m_3} + \frac{4}{3} \frac{\beta_3}{m_3} p_3^3
\end{equation}
\begin{equation}
\label{eq:force_3_triatomic}
\frac{d p_3}{dt} = -k_3( x_3 - x_2) .
\end{equation}
We can add Eqs. \ref{eq:force_1_triatomic}, \ref{eq:force_2_triatomic}, and \ref{eq:force_3_triatomic} to show that $\frac{d}{dt} (p_1 + p_2 + p_3) = 0$. Again, we find that the total pseudomomentum $p_1 + p_2 + p_3$ is a constant of motion
\begin{equation}
\label{eq:total_momentum_triatomic}
p_T = p_1 + p_2 + p_3 .
\end{equation}
We can therefore use Eq. \ref{eq:total_momentum_triatomic} to eliminate one of the $p_i$ in the equations. 

We now specialize to the case of the symmetrical linear triatomic chain where
\begin{equation}
\label{eq:symm_triatomic}
\begin{cases}
m_1 = m_3 = m \ \ \ \ \ \ m_2 = M \ \ \ \ \ \ k_1 = k_3 = k \\
\beta_1 = \beta_3 = \beta_m \ \ \ \ \ \ \beta_2 = \beta_M 
\end{cases} .
\end{equation}
In the case without minimal length, it is known that this dynamical system exhibits two longitudinal modes characterized by anti-symmetric ($x_1 = - x_3$, $x_2 = 0$) and symmetric ($x_1 = x_3 - L$, $x_2 = -\frac{2m}{M} x_1$) behaviour. We shall examine the effect of minimal length on these two modes.

For the anti-symmetric mode $x_2 = 0$. Therefore $\frac{dx_2}{dt} = 0$ and from Eq. \ref{eq:vel_2_diatomic} we conclude that
\begin{equation}
\label{eq:p_2_triatomic_antisym}
p_2 = 0 .
\end{equation}
Correspondingly, the momentum of the central atom will be zero. From $x_1 = -x_3$, we can show using Eqs. \ref{eq:vel_1_triatomic} and \ref{eq:vel_3_triatomic} that 
\begin{equation}
\label{eq:p_1_p_3_triatomic_antisym}
p_1 = -p_3 .
\end{equation}
Now we can differentiate Eq. \ref{eq:force_1_triatomic}, noting that in this mode $\frac{dx_2}{dt} = 0$, and use Eq. \ref{eq:vel_1_triatomic} to obtain a Duffing equation
\begin{equation}
\label{eq:p_1_triatomic_antisym}
\frac{d^2 p_1}{dt^2} + \frac{k}{m} p_1 + \frac{4}{3} \frac{k}{m} \beta_m p_1^3 = 0 .
\end{equation}
We can solve this with the initial conditions
\begin{equation}
\label{eq:initial_conditions_p_1_triatomic_antisym}
p_1|_{t = 0} = p_{10} \ \ \ \ \ \ \frac{dp_1}{dt} \bigg|_{t = 0}  = -k x_{10} 
\end{equation}
where $x_{i0} = x_i |_{t = 0}$ in terms of a Jacobi elliptic function (Eq. \ref{eq:momentum_ho}). We can then plug back the pseudomomentum $p_1$ to Eq. \ref{eq:vel_1_triatomic} and perform the integration to completely solve for the dynamical behaviour of the mode as effected by the presence of minimal length. Based on Eq. \ref{eq:p_1_triatomic_antisym} it turns out that the effect of minimal length on the anti-symmetric mode is to modify the eigenfrequency to the value $\omega_A$ given by
\begin{equation}
\label{eq:omega_triatomic_antisym}
\omega_{A}^2 = \frac{\pi^2 \frac{k}{m} \bigg( 1 + \frac{4}{3} \beta_m A_A^2 \bigg) }{4 \textrm{K}^2(\mu)}
\end{equation}
where
\begin{equation}
\label{eq:amp_triatomic_antisym}
A_A = \sqrt{\frac{3}{4 \beta_m}}  \sqrt{ \sqrt{1 + \frac{8 \beta_m }{3} \bigg( m k x_{10}^2 + p_{10}^2 + \frac{2}{3} \beta_m p_{10}^4   \bigg)}   - 1}
\end{equation}
\begin{equation}
\label{eq:modulus_triatomic_antisym}
\mu = \frac{2 \beta_m A_A^2}{ 3 + 4 \beta_m A^2} .
\end{equation}
The energy of the mode is increased by the existence of a minimal length since $\omega_A > \sqrt{\frac{k}{m}}$.

For the symmetric mode we have $x_1 = x_3 - L$. Using Eqs. \ref{eq:vel_1_triatomic} and \ref{eq:vel_3_triatomic} we find that
\begin{equation}
\label{eq:p_1_p_3_triatomic_sym}
p_1 = p_3 .
\end{equation}
Also, we have in this mode $x_2 = -\frac{2m}{M} x_1$. With Eqs. \ref{eq:vel_1_triatomic} and \ref{eq:vel_2_triatomic} we obtain
\begin{equation}
\label{eq:p_1_p_2_triatomic_sym}
p_1 = - \frac{1}{2} p_2
\end{equation}
and
\begin{equation}
\label{eq:gup_constraint_triatomic_sym}
\beta_m = 4 \beta_M .
\end{equation}
We note that Eq. \ref{eq:gup_constraint_triatomic_sym} is a necessary requirement for the mode to have a solution. By differentiating Eq. \ref{eq:force_1_triatomic} and using Eqs. \ref{eq:vel_1_triatomic} we can obtain the Duffing equation
\begin{equation}
\label{eq:p_1_triatomic_sym}
\frac{d^2 p_1}{dt^2} + \frac{k}{M} \bigg( \frac{2m + M}{m}  \bigg) p_1 + \frac{4}{3} \frac{k}{M} \bigg( \frac{2m + M}{m}  \bigg) \beta_m p_1^3 = 0 .  
\end{equation}
Comparing the form of the Duffing equation above to those we have encountered previously readily shows that there is an increase in the eigenfrequency of the mode.

\section{Conclusion}
\label{sec:conclusion}

We obtained exact analytical solutions to the classical equations of motion and analyze the dynamical implications due to the existence of a minimal length for the free particle, particle in a linear potential, anti-symmetric constant force oscillator, harmonic oscillator, vertical harmonic oscillator, linear diatomic chain, and linear triatomic chain. It turns out that a free particle will move faster and that a particle in a linear potential will fall faster to the region of lower potential when there is a minimal length. For the oscillator systems that we considered (anti-symmetric constant force oscillator, harmonic oscillator, vertical harmonic oscillator, linear diatomic chain, and linear triatomic chain) we observed one common feature, an increase in the characteristic frequency of the system.

%\begin{acknowledgements}
%\end{acknowledgements}

% BibTeX users please use
\bibliographystyle{spbasic}
\bibliography{references}   % name your BibTeX data base

%\begin{thebibliography}{3}
%\bibitem[Author I(1999)]{Ref1}
%Author I (year) Article title. Journal Title-Abbreviated %Vol: pp--pp
%\bibitem[Author and Smith(2001)]{Ref2}
%Author I, Smith J (year) Book title. Publisher, Place, pp numbers
% Format for proceedings
%\bibitem[Author and Smith(2003)]{Ref3}
%Author I, Smith J (year) Paper title. In: Editor, A. (ed.) Proceedings
%Title, Location, Date, pages. Publisher, Place
% etc
%\end{thebibliography}

\end{document}